\documentclass[%
 reprint, 
 superscriptaddress,
 amsmath,amssymb,
 aps,
]{revtex4-2}

\usepackage{graphicx}
\usepackage{dcolumn}
\usepackage{bm}
\usepackage[mathlines]{lineno}

\begin{document}

\preprint{APS/123-QED}

\title{Paste Extrusion Generates a Surface Lubrication Layer.}

\author{Richard T. Benders}
\affiliation{Physical Chemistry and Soft Matter, Wageningen University \& Research, The Netherlands}
\author{Menno Thomas}
\affiliation{Zetadec, Nudepark 73A, Wageningen, The Netherlands}
\affiliation{Animal Nutrition,  Wageningen University \& Research, The Netherlands}
\author{Thomas M.M. Bastiaansen}
\affiliation{Animal Nutrition,  Wageningen University \& Research, The Netherlands}
\author{Raoul Fix}
\affiliation{Physical Chemistry and Soft Matter, Wageningen University \& Research, The Netherlands}
\author{Mario Scheel}
\affiliation{Synchrotron SOLEIL, Saint-Aubin, France}
\author{Guido Bosch}
\affiliation{Animal Nutrition,  Wageningen University \& Research, The Netherlands}
\author{Sonja de Vries}
\affiliation{Animal Nutrition,  Wageningen University \& Research, The Netherlands}
\author{Jasper van der Gucht}
\affiliation{Physical Chemistry and Soft Matter, Wageningen University \& Research, The Netherlands}
\author{Joshua A. Dijksman}
\affiliation{Van der Waals-Zeeman Institute, Institute of Physics, University of Amsterdam, The Netherlands}
\affiliation{Physical Chemistry and Soft Matter, Wageningen University \& Research, The Netherlands}

\date{\today}

\begin{abstract}
Dense particle-fluid mixtures, or \emph{pastes}, are encountered in the production of various materials, including animal feed~\cite{Thomas1997}, human food~\cite{VanderSman2023}, pharmaceuticals~\cite{Fitzpatrick2007}, and biomass for bioenergy~\cite{Stelte2012}. The flow behavior of such dense deformable particulate media is poorly understood, as the interplay between applied stresses, particle deformability and interstitial fluids can be very complex. One challenging context is high pressure pipe flow, encountered in extrusion~\cite{Benbow1993,Bouvier2014}. Despite its widespread use, many questions remain about how during high pressure flow of the paste, the particle-fluid mixture behaves and interacts with boundaries~\cite{Barnes1995,Kalyon2005,Coussot2005}. We show how high pressure paste extrusion induces the formation of a fluid boundary thinner than the particle size~\cite{Lyu2021}. The induced fluid layer emerges from a pressure-induced phase segregation process~\cite{Benbow1987,Coussot2005,Pan2023}. The fluid layer is sufficiently thin to affect particle-wall contacts, making the paste friction coefficient tunable. Our results so offer potential pathways for reducing energy consumption and even extrusion product composition and failure~\cite{Mannschatz2010}.

\end{abstract}

\maketitle


\section*{Introduction}
The flow behavior of concentrated suspensions and pastes plays a role in processes ranging from human food production and additive manufacturing of plastics and concrete structures to medical and clinical applications~\cite{Benbow1987,rabideau2010extrusion,Mascia2006,Salmasi2023,Habib2008,Demir-Oguz2023,Pan2023,Rough2000,VanderSman2023,Ness2022}. During the extrusion of these mixtures of particles and a small amount of liquid, the two-phase mixture is pressured into a pre-shaped channel, usually called the \emph{die}. High extrusion pressures are required to promote particle bonding in the extrudate~\cite{Pietsch2001} and produce a rigid end product. Despite the widespread use of pastes, the flow behavior of these materials still remains poorly understood. Due to the pressures involved, the flow of dense pastes is often characterized as plug flow~\cite{Barnes1995,Kalyon2005}, referring to the perspective that the dense particle phase moves through the die as a rigid body. Yet, these high-pressures conditions in extrusion complicate paste flow substantially: the high pressure flow environments are challenging to probe experimentally, but the elevated stress levels also fundamentally changes the paste flow dynamics. 

In particular, for extrusion context, applied pressures can be high enough to induce particle deformations. The particle deformations make the pore space between the particles change substantially over time and space during the flow. The liquid is present in the pore space, but its role in determining the overall flow dynamics of the paste is poorly understood, as most suspension physics focuses on dynamics in the particle phase~\cite{Ness2022}. High pressure paste flows are hence fundamentally different from dense non-deforming particle suspension flows. This makes our microscopic understanding of the typically non-Newtonian flow behavior of pastes very limited~\cite{Barnes1995,Kalyon2005}. Another challenge is the numerous tuning parameters involved \cite{Benbow1993,Bouvier2014}; lacking a framework of understanding paste flows also limits our ability to understand the interrelatedness of these parameters. 

\begin{figure}[!th]
\centering
\includegraphics[width=0.48\textwidth]{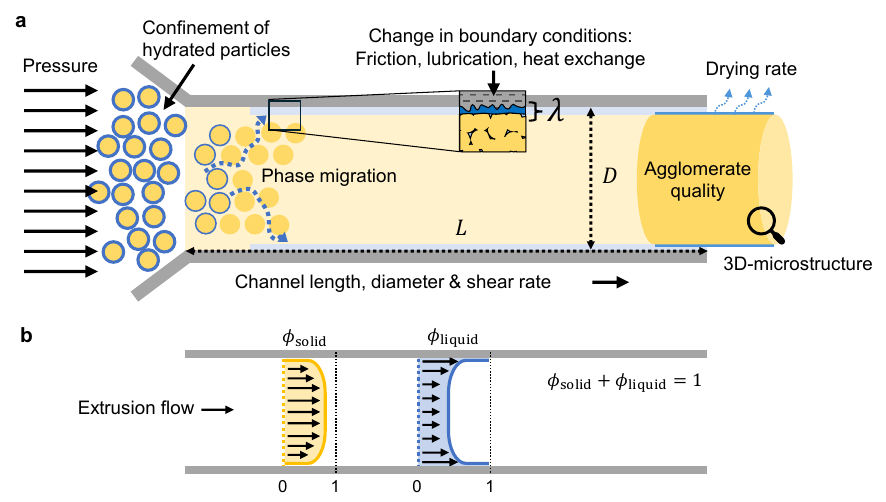}
\caption{ (a) Sketch of the high pressure extrusion flow dynamics. A steam-wetted powder is pressed into a die channel. During the transport of the paste through the channel, the paste is consolidated and its liquid content migrates towards the boundary between the consolidated plug and the die wall, affecting local friction and heat exchange and final agglomerate behavior. Experimental control parameters are channel length $L$, diameter $D$ and flux $Q$. We quantify the lubrication layer thickness $\lambda$ present on the extrudate via its drying process. (b) During channel flow, the density profile of the solid $\phi_{\rm solid}$ and the liquid $\phi_{\rm liquid}$ change from homogeneous to highly depleted/concentrated at the wall ($\phi_{\rm solid},\phi_{\rm liquid}$ respectively) while the overall compressibility of ingredients is limited ($\phi_{\rm solid}+\phi_{\rm liquid}=1$).  
}\label{fig:main:schematic}
\end{figure}

Here we show that paste extrusion flow dynamics is associated with the formation of a liquid film on the surface of the paste. The film is thinner than the particle size~\cite{Lyu2021} and emerges from the paste pore space in a phase segregation process. We indicate the main phenomenology in Fig.~\ref{fig:main:schematic}a, along with our experimental control parameters. The liquid film is only micrometers thick, and located between the extrusion die and the extrudate. The frictional role of the film determines the mechanical energy usage and temperature evolution during the extrusion process, thereby also controlling the binding process between the particles and hence a large number of product quality parameters, which we describe elsewhere \cite{benders2025a,benders2025b}. Optimization of the material and factory performance of extrusion processes of course happens already on a daily basis via paste ingredient adjustments, die shape modification and process temperature and pressure alteration~\cite{Karkania2012,Bouvier2014}. Our results allow to quantitatively connect these parameters via knowledge of the dynamics of the lubrication layer, and can so have major implications for many industrial processes. 

\subsection*{Phase segregation}
During paste extrusion, physical forces act on the solid-liquid mixture, deforming the particles and causing the solid particles and liquid fraction to undergo relative displacement. These phenomena often referred to as ``migration'', ``demixing'' or solid-liquid \emph{phase separation}. The phase migration effect is schematically indicated in Fig.~\ref{fig:main:schematic}b, highlighting how the relative local fractions of solid and liquid $\phi_{\rm solid},\phi_{\rm liquid}$ change from uniform at the channel entrance to highly localized ($\phi_{\rm liquid}$) at the end of the channel. In general, such solid-liquid separation is known to result in flow blockages, product inhomogeneity and product failure~\cite{Habib2008,Gotz1993,Yu1999,Mannschatz2010}. It was conjectured long ago that there is also a beneficial effect: the liquid-rich layer can form between the wall of the die channel and the flowing paste \cite{Benbow1987}, resulting in self-lubricating paste flow. Such liquid lubrication layers between rigid surfaces are known to change the surface sliding resistance by orders of magnitude~\cite{popov2010contact, Wasche2014}, particularly in the so-called ``mixed''-regime where a lubrication layer starts to reduce the solid-to-solid sliding friction \cite{Taylor2022}. As the lubrication layer affects friction or paste \emph{tribology}, the layer influences the mechanical energy use and heat generated inside the extrusion channel at the (sub)micrometer length scale. In the extrusion die, precise tuning of both energy and temperature are necessary to induce and control agglomeration of the particulate mixture through various physical and chemical transitions, such as chemical reactions, melt transitions, and sintering \cite{Pietsch2001}. Control over these transitions is crucial in the formation of a high-quality end product \cite{Pietsch2001,Bouvier2014,Thomas1997,Thomas1998,VanderSman2023}. The control over liquid surface migration induced lubrication is thus an essential step in paste extrusion. Our \emph{post-extrusion surface liquid evaporation} measurements quantify this phase migration by measuring the lubrication film thickness, allowing for deep insight into the flow dynamics of pastes under pressure.

\begin{figure*}[!bth]
\centering
\includegraphics[width=1\textwidth]{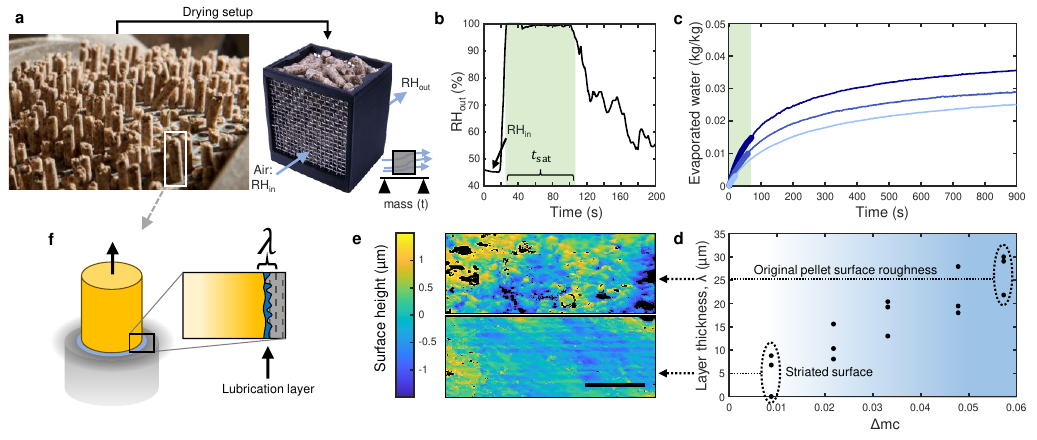}
\caption{(a) Cylindrically shaped pellets extruded through holes within a metal ring-die. The thickness of the lubrication layer is determined immediately after extrusion through a controlled evaporation measurement on freshly sampled pellets, recording sample mass and relative air humidity ($\mathrm{RH}$). (b) Evaporation measurements show an initial saturation period ($\mathrm{RH}>95$\%, denoted as $t_{sat}$) due to evaporation of surface water. (c) The mass-detected average amount of evaporated water depends on the amount of steam added during the conditioning step: the change in water content ($\Delta \mathrm{mc}$) of $0.057$ (top), $0.033$ (middle), and $0.009$ (bottom) in kg water per kg ingredients. Bold lines: the amount of water that evaporates during $t_{\mathrm{sat}}$ as indicated in (b), quantifying the amount of surface water that evaporates. (d) Total surface water evaporation converted to layer thickness $\lambda$, which increases as more water is added in the steam conditioning process ($\Delta \mathrm{mc}$). (e) Surface topology scans of dried pellets do not show striations along the extrusion direction when the lubrication layer is thicker than the intrinsic height of the asperities in the metal die (top). A thinner lubrication layer thickness results in clear striations on the pellets surface along the extrusion direction—running left to right (bottom). Scale bar: 50 $\mathrm{\mu m}$. These images confirm that the metal die and the pellet are largely separated by the lubrication layer, schematically represented in (f). }\label{fig:main}
\end{figure*}

The lubrication layer thickness was suggested to be a layer too thin to measure directly in any realistic extrusion process \cite{Benbow1987,Barnes1995,Kalyon2005,VanderSman2023}. It has thus so far been impossible to verify its existence or to understand how this lubrication layer emerges and depends on ingredients and process parameters. Various experimental methods have been developed to analyze liquid phase migration in extrusion processes at laboratory scale \cite{Benbow1987,Benbow1993,Tomer1999,Tomer1999a,Wildman1999,Gotz2003,Bonn2008}. These studies revealed that flow behaviour during extrusion is highly dependent on ingredient and process variables, including solid-liquid mixing ratio, particle size, deformability of particles, flow rate, pressure, and viscosity \cite{Mascia2006,Benbow1993,Martin2006}, yet stopped short of directly showing the existence of such a lubrication layer, let alone quantify it. 
Our surface liquid evaporation measurements \cite{Benders2022}, directly reveal (1) that this lubrication layer exists, (2) that it is only several micrometers thick, and (3) that the thickness is set by process parameters. We (4) confirm that lubrication tuning results in frictional changes by observing energy savings and temperature modulation. We demonstrate our results via pilot-scale pelleting plant trials, with realistic ingredient streams as typically used in biofuel and animal feed production. Our results can be translated to similar unit operations used in ceramics, foods, and pharmaceuticals production.

\section*{Experimental approach}
Pellets are formed in a unit operation known as the \textit{pellet extrusion process} (Fig.~\ref{fig:main:schematic}a), where particle mixtures, often from multiple ingredients, are pressed by a roller through cylindrically shaped holes in a metal ring-die --- see Fig.~\ref{fig:main}a. The pellet extrusion process transforms the loose and low-density particle mixture into densely packed solid pellets.

Unlike hard inorganic particle pastes used in e.g., concrete extrusion, organic plant-based particles used in food and feedstuff production exhibit hygroscopic properties, are capable of binding or releasing water, and undergo deformation and physico-chemical transitions while being processed. Consequently, the interaction mechanism and physical forces at the boundary layer between the metal extrusion die and the food/feed paste \cite{VanderSman2023} are expected to differ from those of hard particles which for example are subject to excluded volume effects at the die-wall \cite{Barnes1995,Benbow1993,Kalyon2005,Chandler2002}. 

\subsection*{Paste generation}
Prior to pelletization, conditioning uniformly mixes the ingredient particles and steam. The steam condenses upon cooling, transferring its latent heat into the particle mix and forms water \cite{Thomas1997,Thomas2020,Skoch1981}. Consequently, the steam conditioning process sets the temperature and amount of water present in the paste before it enters the extrusion die. It is known that increasing the water content in the mixture lowers the mechanical energy required during pellet extrusion \cite{Skoch1981}.
While the water-induced lubrication in pellet extrusion is evident, the underlying mechanism remains unclear. The observed layer is too thin to be related to a particle migration phenomena, although such effects can also significantly change effective paste viscosity~\cite{coussot2007rheophysics,HUANG_BONN_2007}. Hitherto, one main assumption is that the paste during pellet extrusion internally retains its uniform moisture distribution $\phi_{\rm liquid}$ (see Fig.~\ref{fig:main:schematic}b) as it had right after the mixing/conditioning step \cite{Maier1992,Lambert2018}. We show that this is not the case: the lubricating effects of added water is caused by the formation of a thin lubricating film that partially fills the pores between the metal die and the sliding pellet (Fig.~\ref{fig:main}f), reducing the real contact area between those respective surfaces, as expected in the mixed-lubrication regime \cite{Basu2014}. Our perspective also explains the observed reduction in equipment wear with added water \cite{Skoch1981}. 

\subsection*{$\lambda$: Lubrication layer quantified}
We provide evidence for the existence of the liquid surface layer through a pilot-scale pelleting trial. Approximately 18 tons of pellets were produced while systematically adjusting the conditioning settings, as described in the Appendix \ref{sec:methods}. The fresh pellets were sampled immediately after extrusion and placed in a custom designed sample container permitting drying of the product (Fig. \ref{fig:main}a). The drying method involves exposing a sample container of pellets to a continuous, known flux of fresh air and recording the inlet and outlet air temperature and relative humidity (RH), along with the sample mass \cite{Benders2022}. In 14 out of 15 drying experiments, a saturation period was observed ($t_{\mathrm{sat}}$), during which the passing air reached saturation (Fig.~\ref{fig:main}b). Over this period, up to 40\% of the water introduced by conditioning was lost through evaporation (Fig.~\ref{fig:main}c), corresponding to the moisture ``flash-off'' commonly observed in pelleting factories \cite{Thomas1997}.

Evaporation of the surface layer thus causes the initial saturation of drying air (indicated in Fig.~\ref{fig:main}b). The amount of water evaporating during this period can be quantified as we know the shape and thus total surface area of the pellets in the drying experiment, and the drying mass and density. We so obtain the average water layer thickness, $\lambda$, on the pellets surfaces (Fig.~\ref{fig:main}d). For the full calculation, see the appendix. The thickness of the lubrication layer increases from a few micrometers, similar to the height of surface asperities on the metal die (Fig.~\ref{fig:main}e) in which case we expect frictional interactions that are also known as Stribeck boundary lubrication~\cite{popov2010contact}. As we increase water content of the paste ($\Delta \mathrm{mc}$) by increasing the steam flow in the conditioning process, the layer thickness increases to several tens of micrometers. Compared to the smallest individual particles produced by the milling procedure, this layer remains very thin \cite{Lyu2021} (Fig. \ref{fig:main}d).

\subsection*{Lubrication affects extrudate surface structure}
From the thickness measurements, we understand that the water layer at the surface between the plug and die wall only partially separates the pellet from the die. Increasing the water transfer during steam conditioning reduces, the area of direct contact between the pellet and die surface, leading to less damage to the particles embedded in the pellet surface. We see evidence for this in Fig.~\ref{fig:main}e\&f and to less wear in the die. More evidence is presented in the Appendix: \ref{ED_surface_topology}, Fig. \ref{ED_fig5}. As the lubrication layer grows, we so expect to enter the ``mixed regime'' lubrication in the Stribeck perspective, and a concomitant reduction in friction. 

The presence of the lubrication layer also induces plug flow in the compressed biomass as there is no visible internal deformation, uniform density and negligible porosity (see supportive X-ray tomography evidence described in Appendix: \ref{ED_microstructure}). 

These collective findings suggest that the lubrication layer in organic particulate materials may not arise from factors like particle size distribution and the excluded volume effect, as observed in hard particle pastes. Instead, it appears to be influenced by the roughness of the die wall. In this scenario, water flows into the die's surface cavities and reduces the shearing of the particles at the paste-die-wall interface by occupying part of the cavity volume, thus making it inaccessible for the feed particles. Consequently, the real contact area between the paste and metal die reduces, and the energy use decreases, as shearing of the water layer is much more favorable than shearing of the feedstuff material along the rough metal wall. This hypothesis aligns with industrial findings indicating that new pellet dies require more energy during pellet extrusion than used dies \cite{Whittaker2017}. We attribute this energy use reduction over time to a decrease in the paste-die-wall contact area as the die's largest asperities smooth out through wear. By smoothing of the surface, the water film can cover a larger area (the ratio of $\lambda$ over the wall roughness increases), further reducing the real contact area between the paste and die \cite{Wasche2014}.

The ability to measure and monitor the lubrication layer thickness enables us to decouple process dependent parameters such as steam use, die-hole dimensions and production rate (the latter, together with die-geometry specifications, determines residence time) from ingredient mixture dependent parameters, such as its chemical composition, particle size, particle size distribution and thermomechanical properties \cite{Bastiaansen2023}. We identify two important consequences: we can now identify the source of mechanical energy use of pellet extrusion in the friction at the paste-die-wall interface, and consequently reduce it. Second, by tuning the lubrication layer, we can adjust the amount of heat added during extrusion, giving control over product treatment and quality.

\begin{figure}[!th]%
\centering
\includegraphics[width=0.45\textwidth]{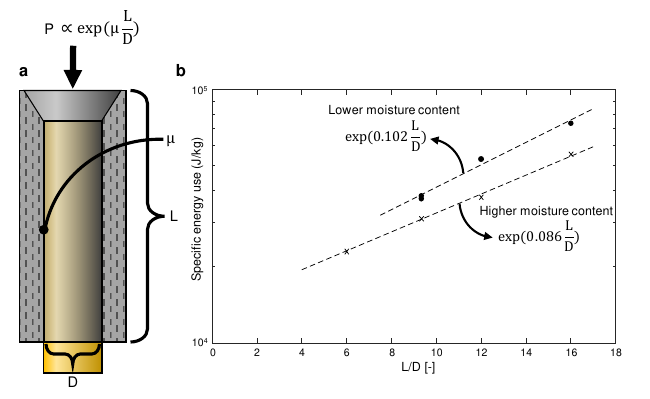}
\caption{(a) During pellet extrusion the specific mechanical energy usage is exponentially proportional to the coefficient of friction and compression ratio of the pelleting die. (b) Increasing the steam use during steam conditioning from a low moisture content ($0.035$ kg steam per kg ingredients) to a higher moisture content ($0.053$ kg steam per kg ingredients), results in a lower energy usage due to a decrease in the amount of friction. This is consistent with previous observations \cite{Skoch1981,Thomas1997}.}\label{ED_fig_LD-ratio}
\end{figure}

\begin{figure*}[!th]%
\centering
\includegraphics[width=.95\textwidth]{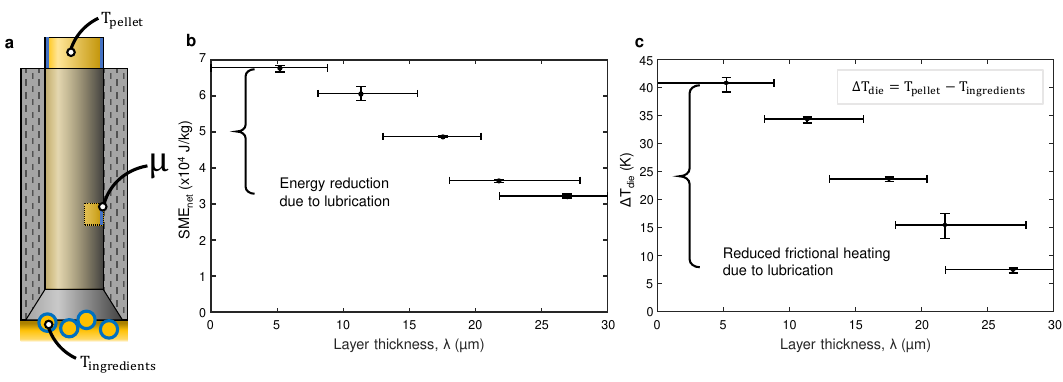}
\caption{(a) Lubrication reduces the apparent die wall friction $\mu$ and, therefore, (b) reduces the net specific mechanical energy $(\mathrm{SME_{net}})$ use with increasing lubrication layer thickness $\lambda$. The reduction in mechanical energy use in particular (c) reduces die heating through friction and consequently affects the change in pellet temperature $\Delta T_{\rm die}$ over the die—the difference between the pellet's temperature after extrusion and the ingredients temperature before extrusion, as well as final pellet temperature. The horizontal error-bars in (b) and (c) correspond to the minimum and maximum values for $\lambda$ as represented in Fig.~\ref{fig:main}d. The vertical error bars represent minimum-maximum range of at least three samples during experimental runs (see Appendix \ref{sec:methods}).
}\label{fig:results}
\end{figure*}

\subsection*{Friction dominates extrudate flow dissipation}
So far, we have only provided evidence of the existence of a liquid surface layer on the extrudate. We can confirm the lubricating effect of this layer by first establishing that energy use in the extrusion of pastes is set predominantly by friction of the extrudate inside the die channel. As pelleting trial do not allow us to change the applied pressure that drives extrusion, we verify the importance of friction in extrusion by probing how mechanical energy dissipation depends on the die geometry. Relying on the same physical principles as the Hagen-Janssen effect~\cite{Tighe2007}, Holm \textit{et al.}~\cite{Holm2006, Holm2011}, derived a relationship that couples compression ratio - die length over die diameter, $L/D$ (-), and the coefficient of friction, $\mathrm{\mu}$, to the pressure required for extrusion during pelletization:

\begin{equation}\label{equation_3}
    P_x(x)= \frac{P_{0}}{\nu} \exp\left(4\mu\nu\frac{L}{D}-1\right)
\end{equation}

In which the extrusion pressure, $P_x$ $(\mathrm{Pa})$, is determined by a prestressing constant, $P_0$ $(\mathrm{Pa})$, the radial-to-axial stress ratio constant, or lateral pressure coefficient, $\nu$ (-), the coefficient of friction and the compression ratio of the die (see~\cite{Holm2006,Sun2023} for further reference). By taking $\nu$ as an ingredient specific constant during pelletization, the pressure required for extrusion increases exponentially with $\mu$ and $L/D$. Furthermore, the extrusion pressure $(\mathrm{N m^{-2}})$ and power consumption $(\mathrm{N\ ms^{-1}=W})$ are related through the volume flow rate $(\mathrm{m^3s^{-1}})$. Consequently, the power consumption of the pellet press, as well as the specific mechanical energy usage - the energy per unit mass - is expected to increase exponentially as a function of $\mu$ and $L/D$ as well, which we confirm in Fig. \ref{ED_fig_LD-ratio}.

Increasing the steam use during steam conditioning, achieved by controlling the temperature of the ingredient mixture after conditioning, leads to higher moisture content in the ingredient mixture and, a larger $\lambda$ and as a result, reduces the mechanical energy demand during pellet extrusion \cite{Skoch1981,Thomas1997}. Our experimental data indicates that increasing steam use from $0.035$ kg\ steam per kg\ ingredients to $0.053$ kg steam per kg ingredients results in a reduction of the apparent coefficient of friction by approximately $16\%$ (see Fig. \ref{ED_fig_LD-ratio}b). The thickness of the lubricating film is directly related to the increased separation of the two sliding surfaces (Fig \ref{fig:main}d\&e). By increasing the lubrication layer thickness the direct contact area between the metal die and the sliding pellet reduces and the relative contribution of $\mu_{\mathrm{solid}}$ with respect to the apparent friction coefficient, $\mu$, decreases, reducing the extrusion sliding resistance. The contribution of the friction within the water film due to viscous damping, is negligible despite the appreciable shear rate achieved (see appendix \ref{appendix:EnergyDissipation}). Hence our experimental observation supports the hypothesis that the increasing use of steam, and consequently the increase in moisture content, contributes to lowering the coefficient of friction between the metal die and the sliding pellet, through the formation of a lubrication layer in the mixed lubrication regime. In the next section, we shall see that we can indeed directly link the extrusion energy consumption to the lubrication layer thickness.

\subsection*{Energy reduction due to lubrication}

As we have seen, the sliding friction between the pellet and the metal die contributes to the energy use of the pellet press. This energy use is proportional to the extrusion pressure and increases exponentially with  $L/D$ and with the apparent coefficient of friction between the die and the paste. The apparent coefficient depends on the solid friction coefficient and is determined by the paste composition~\cite{Stasiak2020,Ibrahim2008}. This dry coefficient of friction can not be actively manipulated during pellet extrusion, yet the formation of a thin lubrication layer can contribute to a smooth transition from the full contact ``boundary lubrication regime'' into the so-called ``mixed regime'' \cite{Wasche2014}, in which part of the solid contact is replaced by a lubrication layer, significantly reducing the coefficient of friction (see appendix \ref{ED_surface_topology}). 

The decrease in the apparent coefficient of friction, $\mu$, at the die-wall surface has a direct impact on the specific mechanical energy (SME) use of the pellet press—the amount of mechanical energy required to produce one kilogram of pellets. This is exemplified in Fig.~\ref{fig:results}b, which shows that increasing the lubrication layer thickness, due to an increase in the initial $\phi_{\rm liquid}/\phi_{\rm solid}$ ratio after steam conditioning, reduces the specific mechanical energy (SME) during extrusion by approximately 50\%. Hence, adjusting the lubrication layer through steam addition enables modification of the friction coefficient within a running production line, without altering the ingredient mixture composition or press/die configuration. In addition to the change of the initial $\phi_{\rm liquid}/\phi_{\rm solid}$ ratio, the extrusion rate is another key factor that modulates friction and energy dissipation during pellet extrusion. An increase in extrusion rate reduces frictional resistance and thus lowers SME, in accordance with Stribeck curve expectations~\cite{Wasche2014} (see appendix Fig.~\ref{fig:ED_Q_vs_SME}). 

Overall, these changes at the die-wall boundary affect the amount of heat dissipated during extrusion. An increase in lubrication layer thickness reduces the heating of the pellet during extrusion, here quantified by $\Delta T_{\rm die}$. This effect is in Fig.~\ref{fig:results}c, where the increasing in lubrication layer thickness is linked to a reduction in $\Delta T_{\rm die}$ due to the reduced die-wall friction. These changes in the extrudate's final temperature also affect final product properties, such as ``durability'', as detailed in appendix \ref{ED physical quality}, and elsewhere \cite{benders2025b,benders2025a}. We reason that these temperature fluctuations change the physico-chemical properties of the organic ingredient streams, which may result in gelatinization and melting of starches and denaturation of proteins, which subsequently affect the nutritional quality and availability during consumption \cite{Thomas2020}. Measurement and subsequently tuning of the lubrication layer is thus essential for further optimization of pelleting processes as it enables a more precise control over the mechanical energy use and product temperature compared to the current set of operating variables.

\begin{figure}[!th]%
\centering
\includegraphics[width=0.45\textwidth]{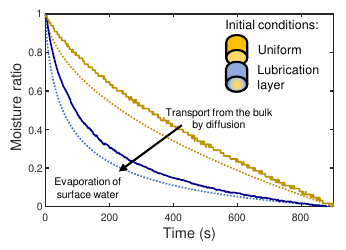}
\caption{Two types of pellets underwent drying using our model drying setup (solid lines), complemented by results obtained through numerical modeling (dashed lines). The results are presented as the pellet moisture ratio, the normalized mass decay (unitless) over $900$ seconds of drying (see Eq. \ref{appB_eq3}). In yellow: experimental drying data of pellets with a uniform moisture content, achieved through vapor sorption (stored for 1 week at over $95\%$ RH), and numerical simulation considering an initially uniform moisture distribution. Blue: experimental drying data of fresh pellets immediately after extrusion and numerical simulation for a pellet with a high moisture content localized at its surface, as schematically depicted. The initial drying rate is strongly affected by the presence of surface water, resulting in a quick drying profile, in contrast to pellets with a uniform moisture distribution, for which the rate of drying is determined by the transport by diffusion. Additional results are shown in Fig. \ref{ED_fig1}.}
\label{fig:numerics}
\end{figure}

\section*{Numerical evidence for $\lambda$}
The concept of liquid migration during extrusion of multi-phase ingredient mixtures is not unique to extrusion and occurs under a wide range of process conditions. Ranging from low viscosity soft matter suspensions, e.g. in food \cite{Nieuwland2023} or human health \cite{Pan2023} applications, as well as during the extrusion of highly viscous flows, like rock melts \cite{Quintanilla-Terminel2019}. As a result of the physical forces, an uneven distribution of the liquid fraction occurs across the product during or after extrusion. However, direct analysis of liquid distribution inside extrusion channels is currently not possible. The higher pressure extrusion conditions make non-invasive techniques like Magnetic Resonance Imaging or Optical Coherence Tomography incompatible or do not have the required spatial resolution~\cite{Tomer1999a,Gotz2003}.

To address these limitations, we also conducted a numerical evaluation of how radial redistribution of water over the pellet influences drying characteristics immediately after extrusion. The numerical results were qualitatively compared with experimental drying data obtained in the experiment. We describe the numerical setup at length in Appendix \ref{extData}. Our modeling demonstrates that the saturation of the drying air and the initial rate of drying, observed during the drying of freshly produced pellets, can only occur if the evaporating water comes directly from the pellet surface. 
This is in contrast to pellets with an elevated but uniform moisture profile, which can be generated through vapor sorption (see Fig. \ref{fig:numerics}; we refer to Fig. \ref{ED_fig1}~b\&c and appendix: \ref{ED_water_distribution} for a detailed description of the numerical model and the results).

\section*{Conclusions}
We confirmed the existence and quantified the thickness of the micrometer-thick self-lubrication layer that forms via particle-fluid phase segregation during the high pressure extrusion of pastes. We show that this lubrication layer directly affects the tribological properties of the paste flow. We used realistic, plant-based materials inside an industrial-scale pellet extruder to evidence how the fundamental aspects of the lubrication layer interlinks industrial process variables. We validated our measurements through numerical simulations of liquid transport around non-uniformly wetted objects, and in doing so, unravel the fundamental mechanism behind a decades-old observation of the lubricating effect of water addition during pellet production~\cite{Skoch1981}.

Our work reveals that the friction of an extrudate within a die is directly influenced by the thickness of the lubrication layer, providing insights for future monitoring of product quality, shape, and performance in food, feed, and pharmaceutical extrusion processes and adjacent extrusion disciplines which process inorganic ingredient streams like ceramics. This work introduces a framework to understand and utilize the role of extrusion parameters such as die geometry, particle size distribution, and ingredient composition. Furthermore, continuous monitoring of the phase segregation process during extrusion presents a novel strategy for process control, with immediate global industrial relevance. For instance, improved lubrication control can yield significant energy savings and enable more sustainable, high-performance extrusion. Real-time tracking of phase segregation may also enhance our understanding of the complex particle-liquid interactions, which we can subsequently use in quality control across a wide range of extrusion-based manufacturing processes. 

\begin{acknowledgments}
We gratefully acknowledge discussions with the ``Pelleting in Circular Agriculture" consortium members. We thank the Wageningen Electron Microscopy Center (WEMC) of Wageningen University \& Research for access to facilities. We thank the Van der Waals-Zeeman Institute for providing access to the Keyence VK-X1000. We thank Zohreh Farmani and Jordi Rijpert for their help during the experimental investigation. We thank Remco Fokkink for helping with the development of software to read out the relevant sensors during the drying experiments. The authors thank Synchrotron SOLEIL for the ANATOMIX beamline time (proposal n° 20220272). ANATOMIX is an Equipment of Excellence (EQUIPEX) funded by the \textit{Investments for the Future} program of the French National Research Agency (ANR), project \textit{NanoimagesX}, grant no. ANR-11-EQPX-0031. This study was financially supported by The VICTAM Foundation, Agrifirm NWE B.V., DSM, Elanco Animal Health, Pelleting Technology Netherlands, Phileo S.I. Lesaffre, Topsector Agri \& Food, Zetadec, and Wageningen University \& Research, as partners in the project  ``Pelleting in Circular Agriculture'' (project number: LWV1965)
\end{acknowledgments}

\section*{Declarations}

\begin{itemize}
\item Conflict of interest/Competing interests: MT owns a company that provides consultancy services for the animal feed industry.
\item Availability of data and materials: all processed data is available in the main article and supplementary materials; raw data will be provided upon reasonable request.
\item Authors' contributions: 
\end{itemize}

\textbf{Richard Benders:} Conceptualization, Methodology, Investigation, Formal analysis, Data Curation, Visualization, Writing - original draft
\textbf{Menno Thomas:} Conceptualization, Methodology, Writing - review \& editing, Supervision, Project administration.
\textbf{Thomas Bastiaansen:} Investigation, Writing - review \& editing.
\textbf{Raoul Fix:} Conceptualization, Methodology, Investigation, Writing - review \& editing.
\textbf{Mario Scheel:} Conceptualization, Methodology, Investigation, Writing - review \& editing.
\textbf{Guido Bosch:} Conceptualization, Writing - review \& editing, Project administration
\textbf{Sonja de Vries:} Conceptualization, Writing - review \& editing
\textbf{Jasper van der Gucht:} Conceptualization, Writing - review \& editing, Formal analysis, Supervision
\textbf{Joshua Dijksman:} Conceptualization, Methodology, Writing - original draft, Writing - review \& editing, Supervision.

\clearpage

\appendix
\counterwithin{figure}{section}

\section{Extended Data}\label{extData}

\begin{figure*}[!th]%
\centering
\includegraphics[width=\textwidth]{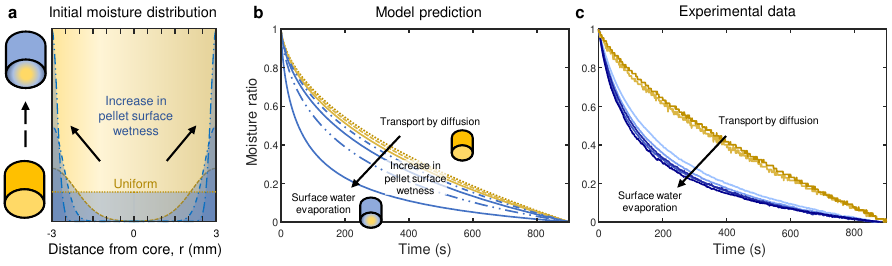}
\caption{(a) A schematic representation illustrates the initial radial moisture distribution used as the initial condition in numerical modeling of the drying process for freshly produced pellets. The moisture ratio (unitless) represents the normalized mass decay over $900$ seconds of drying (see Eq. \ref{appB_eq3}). Numerical simulations provide a qualitative comparison of the effect of radial phase segregation on drying behavior, shown in both numerical simulations (b) and experimental drying data (c) for fresh (blue lines) and rehydrated pellets (yellow lines) using the moisture distributions as the initial condition (a). The difference in drying behavior is caused by radial phase segregation, specifically the formation of a lubrication layer during pellet extrusion.}\label{ED_fig1}
\end{figure*}

We examined the pellet drying behavior, internal and external pellet microstructure, and surface topology of pellets produced under systematically varied process conditions, adjusting only the amount of added steam to the ingredient mixture when formulating the paste, thereby changing its initial water content prior to compaction, $\Delta \mathrm{mc}$. Our objective was to determine whether the self-lubricating liquid surface layer emerges during pellet extrusion. In the processing of pastes and other dense particulate suspensions, such phase segregation processes are also referred to as ``spatial demixing," ``liquid migration," ``phase redistribution," ``drainage," or ``self-filtration" \cite{Pan2023,Martin2006}. According to our hypothesis, a thin lubricating film forms between the pellet and the metal die due to phase segregation (Fig. \ref{fig:main}a), which fills the asperities between the two sliding surfaces and reducing the frictional contact area. As a result of the radial liquid migration, the moisture distribution across the pellets becomes non-uniform, with the exterior surface of the pellet facing the die wall experiencing increased moisture levels during extrusion compared to the pellet's interior volume \cite{Gotz1993}. Through the combined circumstantial evidence presented below, we conclude that such phase migration occurs during pellet extrusion.

While the quantification of the lubrication layer thickness is detailed in the main text, here we additionally demonstrate how adding water in the form of condensed steam reduces the apparent coefficient of friction  within the mixed lubrication regime. We also discuss how tuning the lubrication layer can be performed to produce high-quality pellets with improved energy efficiency, monitoring physical pellet characteristics, and measuring the energy use within the pelleting mill. This involves balancing out the thermal energy required to produce steam and the mechanical energy required during pellet extrusion. 

\subsection{Pellet production}
The extrusion of pellets through individual die-holes occurs semi-continuously. The ring-die rotates at a constant velocity; however, the rollers, fixed in position with respect to the rotating die, periodically pass the openings of the die-holes as the die rotates, causing compression of the ingredient mixture and subsequent extrusion of the material through the die-hole. The compression frequency, representing the number of compression steps per second, and the shear rate of the extrusion process are determined by the mass-flux of the ingredient mixture into the die, the rotational velocity of the die, and the number of rollers. The length of the pellets is controlled by knives fixed at a specified distance outside from the rotating die. Pellets shorter than this specified distance pass underneath the knives until they reach the required length, at which point they are sheared off upon collision with the knives.

\subsection{Water distribution inside pellets: $\phi_{\rm liquid}$}\label{ED_water_distribution}

In our numerical simulations, the initial moisture distribution $\phi_{\rm liquid}$ along the radial coordinate $r$ 
was described by a half-Gaussian function $M(r)$—a Gaussian centered at the pellet-air interface ($r_p$).
\begin{equation}\label{equation_1}
    \phi_{\rm liquid} \equiv M(r) = M_{\rm surf}\exp\left[-\frac{1}{2}\left(\frac{r-r_p}{\sigma}\right)^2\right]
\end{equation}
The Gaussian function allows us to set both the height ($M_{\rm surf}$) representing the amount of water at the air-pellet surface ($\mathrm{mol/L}$), and the width ($\sigma$) of the moisture distribution, thereby controlling the shape of the moisture distribution inside the pellet. It enables us to assure that every simulation was initiated with a constant initial overall moisture content, making sure the integral $\int M r\ dr$ over the domain $r=0$ to $r=r_p$ was constant at the start of a simulation (see appendix \ref{appendix:COLSOM_Model}). Using this function various distributions, e.g. a uniform moisture distribution, weak moisture gradients (resembling experimental results in ref. \cite{Gotz1993}) or strong moisture gradients (resembling experimental results in ref. \cite{Quintanilla-Terminel2019}) could be obtained using the same mathematical expression without violating physical constraints (e.g., preventing water concentrations above those of pure water, $55.5\ \mathrm{mol/L}$) (Fig. \ref{ED_fig1}a). 

At the pellet's surface, water evaporation is driven by the vapor pressure difference between the water exposed on the pellet's surface and the vapor pressure of the air passing through the bed of pellets at a given temperature \cite{Baer1958}. If a water layer is present on the pellet's surface, the maximum theoretical evaporation rate from this surface can be determined using, for example, the Hertz-Knudsen-Schrage equation based on classical kinetic theory or alternatively, equations based on statistical rate theory \cite{Zhang2017}. However, during the drying experiments, we initially observe the saturation of the air passing through the bed of pellets (see Fig. \ref{fig:main}b), leading us to conclude that the evaporation rate from the pellet's surface is not limited by the maximum theoretical evaporation rate but is limited by the vapor capacity of the influx of fresh air. In other words, the initial evaporation of the surface water—the evaporation of the lubrication layer—occurs at such high rates that the passing air reaches its saturation point as long as sufficient water is available on the pellet's surface. When the water layer has evaporated, the rate of evaporation reduces, as indicated by the change in slope in Fig. \ref{fig:results}c.

Consequently, for the numerical model, mass transfer coefficient ($k_c$) and effective diffusion coefficient ($D_{\rm eff}$) were selected in a way that the transport rate at which water moves within the pellet (diffusion of water) is much lower than the rate of evaporation (mass transfer rate representing the change of liquid water on the surface into water vapor within the passing air). This is associated with a mass Biot number ($\mathrm{Bi_m}$) much larger than $1$ (see Appendix \ref{appendix:COLSOM_Model}). With a mass Biot number much larger than $1$, we ensure that the rate of evaporation at the surface does not limit the drying process, as the maximum evaporation rate is lower than the vapor capacity of the passing fresh air, as observed in the experimental data (see Fig. \ref{fig:main}b).

Here, we show that the experimental drying behavior complements the expected numerical behavior when significant phase migration occurs during pellet extrusion.

The results obtained through numerical modelling (Fig. \ref{ED_fig1}b) show how a shift in water distribution, ranging from uniform (yellow) to an increased exposure of surface water (blue), affects the expected drying behaviour. Experimentally and numerically, we were able to distinguish two different cases by qualitative comparison. First, the case in which rehydrated pellets with a uniform moisture content were dried in a control experiment and secondly those of fresh sampled pellets, directly after extrusion (Fig. \ref{ED_fig1}c).

In the first case, a control experiment was performed in which pellets were rehydrated through hygroscopic vapor sorption. The moisture content throughout these pellets can be assumed near-uniform due to the long equilibration time (stored 1 week at \textgreater 95\% RH) \cite{Vego2023} and consequently the drying process resembles that of a finite cylinder with surface evaporation (\cite{Crank1975}). Here, the drying rate is controlled by diffusion through the sample (yellow lines in Fig. \ref{ED_fig1}b and Fig. \ref{ED_fig1}c). During the control experiment, no saturation of the passing air was observed. The same holds true for the simulation, during which the calculated initial rate of evaporation averaged over 10 seconds ($\approx10^{-8}\ \mathrm{kg\ m^{-2}s^{-1}}$) was much smaller than the maximum evaporation rate based on the vapor capacity of fresh air ($\approx10^{-4}\ \mathrm{kg\ m^{-2}s^{-1}}$) flowing through a bed of pellets and along the pellets' surface. The latter was estimated using the vapor capacity of fresh air ($10^{-2}\ \mathrm{kg_{vapor}\ m^{-3}}$), the volumetric flow rate of fresh air into the bed ($10^{-3}\ \mathrm{m^{3}s^{-1}}$), and the total surface area of the pellets within the sample container ($10^{-1}\ \mathrm{m^{2}}$). This indicates that the evaporation rate from a bed of pellets with a uniform moisture content is much smaller than the rate required for saturation of the air. Increasing $D_\mathrm{eff}$ to $10^{-7}\ \mathrm{m^2s^{-1}}$ in the numerical drying model increases the initial evaporation rate up to $\approx5\times 10^{-7}\ \mathrm{kg\ m^{-2}s^{-1}}$, but this still falls short of the required evaporation flux for saturation of the air, despite using a diffusivity value several decades larger than the typical reported experimental values \cite{Panagiotou2004,Maier1992,Lambert2018}. 

In the second case, freshly sampled pellets were sampled and dried directly after extrusion. In contrast to the drying process of pellets with uniform moisture content, freshly sampled pellets exhibit a distinctly different drying behaviour (blue lines in Fig.\ref{ED_fig1}b and Fig. \ref{ED_fig1}c). The initial drying rate for fresh pellets is notably faster during both experimental and computational drying compared to the uniform initial condition driven by diffusion.

The qualitative comparison between the numerical model and the experimental drying data supports the hypothesis that a liquid surface layer emerges during pellets extrusion. Combining the quick initial drying of the fresh pellets together with the observation of air saturation (Fig. \ref{fig:main} b), we conclude that water must be present in close proximity to the pellet exterior surface as a result of liquid surface migration during pellet extrusion. 

\newpage
\subsection{Pellet microstructure}\label{ED_microstructure}
During pellet extrusion, the finely ground organic ingredients are bound into a cylindrical shaped pellet. Such organic ingredient streams deform easily at the typical process pressures, resulting in densification of the ingredient mixture, expelling pore fluid and enhancing the agglomeration process through various bond types \cite{Pietsch2001}. The mechanical properties \cite{Glenn1991,Glenn1992}, as well as the glass and melt transition temperatures of such organic plant derived products are strongly influenced by the local moisture content \cite{Bouvier2014,Thomas2020}. Consequently, the local increase in moisture content due to the presence of a water rich lubrication layer at the pellets surface is expected to affect the surface microstructure of the pellet.
Hence, we employed x-ray tomography at the synchrotron SOLEIL, utilizing the ANATOMIX beamline \cite{Weitkamp2022}, to analyze the pellet microstructure. This analysis focused on identifying density gradients and structural changes associated with shear deformations and moisture gradients. Such gradients and structural changes can offer insights into the flow profile inside the extrusion die and the local moisture content during extrusion. 

\subsubsection{A uniform radial pellet density profile}
At the ANATOMIX beamline, a monochromatic X-ray beam of 16.83 keV from a DCM (double crystal monochromator), at a distance of 20 cm (sample to scintillator; pixel size of 3.07 µm; exposure time 50 ms) was used to analyse the radial density profile of pellets, avoiding beam hardening artifacts. Such artifacts are typically observed in commercial x-ray tomographs, in which hardening of polychromatic x-ray beams results in a concave apparent density profile after image reconstruction. One of our hypothesis was that we expect a density gradient in a narrow region close to the pellets surface. In this region the increased moisture level could plasticise the organic material \cite{Glenn1991}, making it more deformable and consequently a higher density can be obtained during the consolidation process. The use of a monochromatic beam, prevents beam hardening and enables a qualitative analysis of the radial pellet density.

\begin{figure*}[!th]%
\centering
\includegraphics[width=\textwidth]{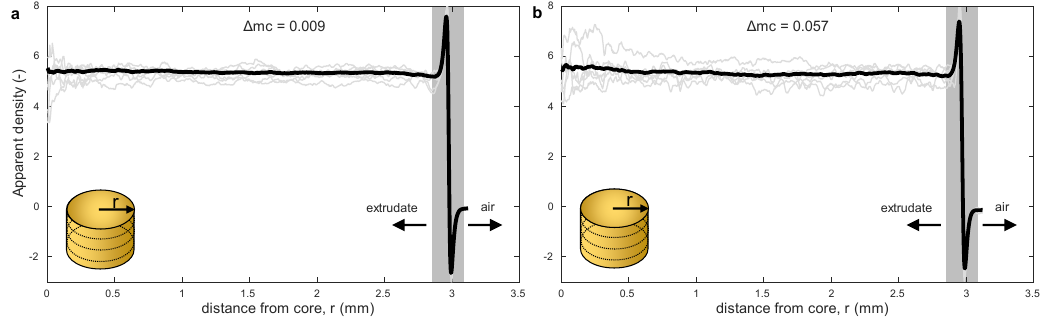}
\caption{Radial density profiles of pellets with a diameter of 6 mm, obtained through analysis of x-ray tomograms obtained at the ANATOMIX beamline, for two process conditions with (a) $\Delta \mathrm{mc}$ $0.009$  (low water content) and (b) $0.057$ (high water content) respectively. The light gray lines in (a) and (b) correspond to individual slices along the 3D volume. The black lines in (a) and (b) correspond to the average density profile over each of the slices within the stack (see Materials \& Methods). The region marked by the gray area, corresponds to the area in which we observe a phase-contrast enhancement artifact. In this region the apparent density could not be determined accurately.}\label{ED_fig2}
\end{figure*}

The apparent density near the surface is subject to phase-contrast enhancement, characterized by a negative apparent density on the exterior (outside) and a positive peak on the interior (inside) of the pellets (gray region Fig.\ref{ED_fig2}). While phase-contrast enhancements are commonly employed to enhance sensitivity, conducting a detailed study of the outer $100 \mathrm{\mu m}$ region near the pellet surface—where we anticipate density changes due to high moisture contents from the lubrication layer—was not feasible due to the phase contrast enhancements. The remainder of the pellet appears to be of uniform density along the radial coordinate, at both high and low moisture levels during production (Fig.\ref{ED_fig2}).

\subsubsection{Surface melt structure}
\begin{figure*}[!th]%
\centering
\includegraphics[width=\textwidth]{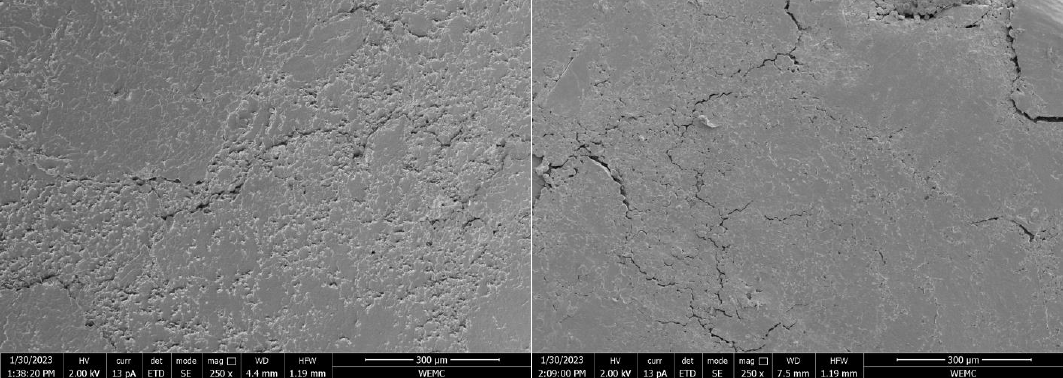}
\caption{Scanning Electron Microscopy images of the pellet's surfaces (left - low water content $\Delta\mathrm{mc} = 0.009$ and right - high water content $\Delta\mathrm{mc} = 0.057$) demonstrate that a network of molten starch granules has formed on the pellet's surface during pellet extrusion. At the pellet's surface, the local moisture content has increased, lowering the melting point temperature of the starch granules to the process temperatures reached during pellet extrusion.}\label{ED_SEM_surface}
\end{figure*}

\begin{figure*}[!th]%
\centering
\includegraphics[width=\textwidth]{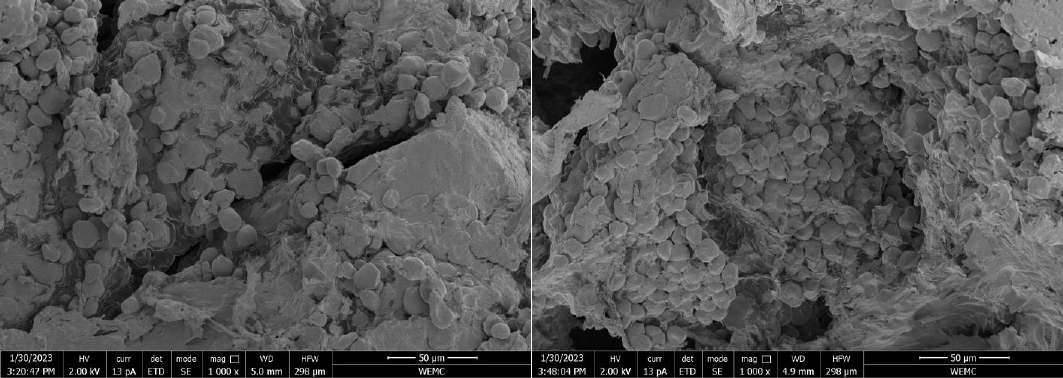}
\caption{Scanning Electron Microscopy images of the pellet's interior (left - low water content $\Delta\mathrm{mc} = 0.009$ and right - high water content $\Delta\mathrm{mc} = 0.057$) demonstrate that only a limited number of starch granules on the inside of the pellet have reached their melting point temperature, while most remain in their crystalline shape. The moisture content on the interior of the pellets is expected to be lower than on the pellet's surface due to liquid phase migration. Consequently, the starch granules on the inside have a higher melting point compared to those at the surface, and only limited melting of the starch occurs during pellet extrusion.}\label{ED_SEM_interior}
\end{figure*}

Complementary scanning electron microscopy (SEM) images of the surface, show the formation of a gelatinized network of starch at the pellets surface (see Fig. \ref{ED_SEM_surface}). Water, absorbed by the organic particles, acts a plasticizing agent and reduces the glass and melt transition temperature of starch and contributes to pellet binding \cite{Thomas2020,Bouvier2014}. Hence, the starch melt is not coincidentally formed at the location where the lubrication layer forms during extrusion, in contrast to the much more limited melt formation on the interior of the pellet (see Fig. \ref{ED_SEM_interior}). The increase in moisture content at the pellet surfaces, acts as a plasticizer and enables a localized lowering of the melt point, effectively glueing the individual particles together at the surface.

\subsubsection{Internal pellet deformation}
\begin{figure*}[!bh]%
\centering
\includegraphics[scale = 1]{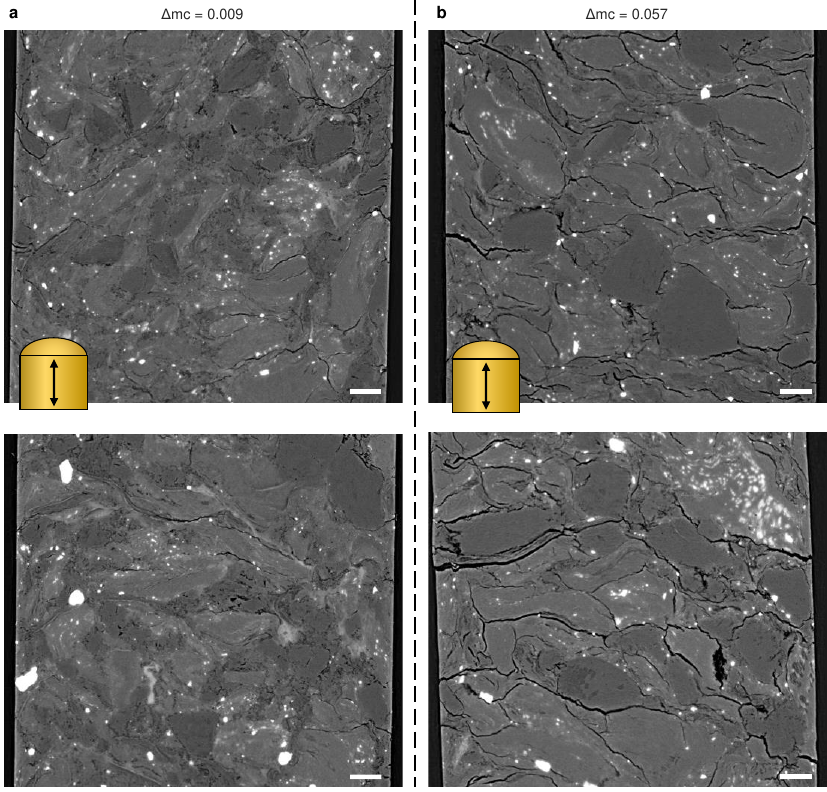}
\caption{Longitudinal cross sections (x-ray tomograms obtained at the ANATOMIX beamline) of four pellets produced during the experimental trial at two different water contents after steam treatment, (a) two at low $\Delta \mathrm{mc}$ and (b) two at high $\Delta \mathrm{mc}$, 0.009 and 0.057 kg water per kg ingredients respectively. No indication for the flow direction is visible within the cross sections. The pellets flow as plugs through the pelleting die \cite{Nielsen2020}. Scale bar indicates $0.5$ cm.}\label{ED_fig4}
\end{figure*}

It has been previously proposed that the ingredient stream moves as a plug inside the die channels  \cite{Nielsen2020}, with a uniform flow velocity through the pellet die. A lubrication layer should facilitate such plug-flow and therefore, signs of internal deformations caused by velocity gradients should be absent. A qualitative analysis of x-ray tomograms confirms this absence of internal deformations along the longitudinal axis as indications of the flow direction (Fig. \ref{ED_fig4}). Additionally, the overall porosity is low (Fig. \ref{ED_fig4}), suggesting that the ingredients are forced into place, approaching unit density during pelletization. Indeed, the estimated process pressures during pelletization \cite{Nielsen2009}, are of similar order, or larger than the typical compressive strength of the particles \cite{Fang2002} and hence the particles are compressed into a tightly packed pellet with a low porosity. These observations support the hypothesis of liquid surface migration, wherein water in between the particles, is expelled from the ingredient matrix into the pellet-die gap during pellet extrusion, leading to the formation of a lubrication layer that enables plug flow inside the pellet die. 

\clearpage
\subsection{Friction, surface topology, and wear}\label{ED_surface_topology}
In rheological experiments, it has been shown that various physical and chemical forces and constraints give rise to so-called slip-flow due to the displacement of the dispersed, solid-like, phase away from the moving boundary. This displacement of the dispersed phase, and hence the enrichment of the low-viscosity, the liquid-like, phase near the boundary, enhances material flow near the moving boundary \cite{Barnes1995}. In the case of stiff particles at the specified process conditions, particle size, excluded volume, and die-wall roughness are crucial properties of the system, determining whether lubrication layers form \cite{Barnes1995}. However, in the case of pelletizing soft and deformable particles of organic origin, parameters such as the excluded volume and mechanisms such as particle diffusion cannot be defined, as the confining pressure creates an essentially uniform density across the flowing zone. Particle size distributions only remain relevant for contact dynamics between the particles and the particle-wall interaction, as they set the intrinsic roughness scale at the paste-die interface. 

Here, we extend the rheological observations to pellet extrusion, reasoning that the water layer between the flowing pellet and metal die, denoted as $\lambda$, reduces the direct contact area between the two sliding surfaces and thus reduces the apparent coefficient of friction. First, we present surface topology scans that show the addition of water during steam conditioning reduces wear on the pellet's exterior surface, attributed to an increasing lubrication layer thickness between the sliding surfaces in the mixed lubrication regime, schematically represented in Fig. \ref{ED_fig5}. Secondly, we demonstrate how the addition of water through steam conditioning reduces the tribology, specifically the apparent coefficient of friction between the surfaces and contributes to lower mechanical energy use during pellet extrusion.

\subsubsection{Surface topology and wear suggest mixed regime lubrication}
\begin{figure*}[!th]%
\centering
\includegraphics[width=\textwidth]{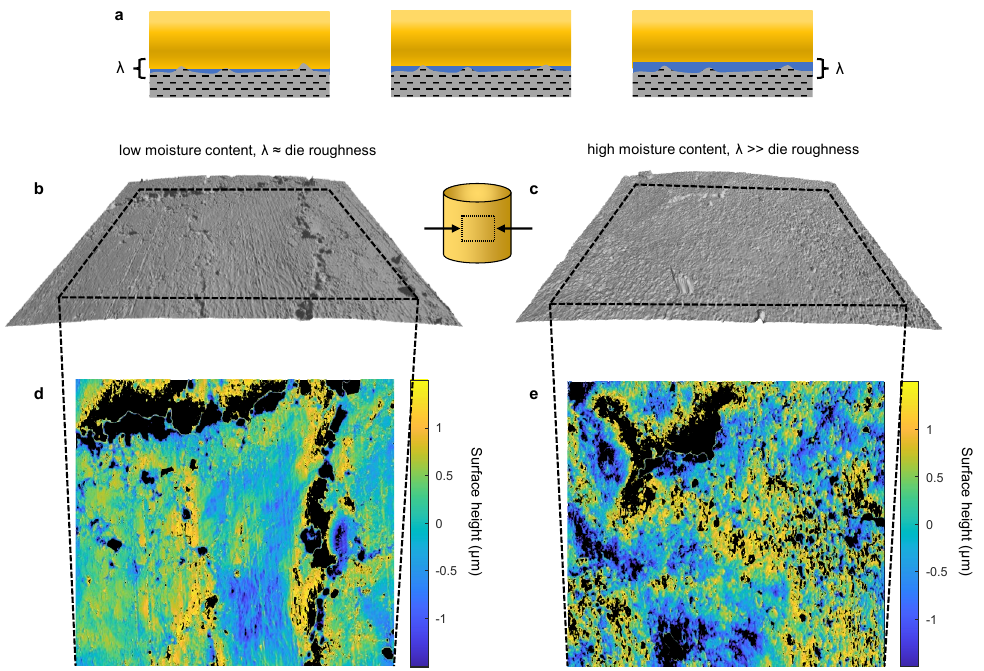}
\caption{(a) Schematic representation of the pellet sliding along the metal surface of the pelleting die with asperities. Increasing lubrication layers are expected to increase the separation between the pellet and the die's surface, reducing their contact area as expected in the mixed lubrication regime \cite{Wasche2014}. (b,c) 3D surface topology scans of two pellets produced at low $\Delta \mathrm{mc}$ and high $\Delta \mathrm{mc}$, $0.009$ and $0.057$ kg water per kg ingredients respectively. (d) \& (e) are the surface height profiles of the pellets scanned in (b) and (c). The black area's correspond to peaks and valleys with a height or depth larger than $1.5 \mathrm{\mu m}$. In (b) and (d) striations along the sliding direction are clearly visible, due to a thin lubrication layer. In (c) and (e) those striations are absent, due to filling of the cavities inside the pellet and the die, resulting in a reduced protrusion of the metal asperities into the extruded pellet. }\label{ED_fig5}
\end{figure*}

Surface topology scans of pellets were conducted, to support the hypothesis of decreased contact area between the two sliding surfaces as a result of a water film with increasing thickness between the pellet and metal surface. Providing evidence for the type of lubrication regime during pellet extrusion. At low water contents, distinct striations are visible along the direction of extrusion (Fig. \ref{ED_fig5}b\&d), indicating a boundary regime lubrication with direct contact between the pellet and die-wall. Under these conditions, the calculated lubrication layer thickness is relatively small in comparison to the expected roughness of the metal die inside the holes (see Fig \ref{fig:main}d), resulting in surface wear due to protrusion of the metal asperities into the softer pellets' surface. We were unable to obtain a measure for the surface roughness of the die during these experiments. This would require cutting of the stainless steel die into measurable pieces and thus destroy valuable equipment. However, the striations disappear when increasing the water content from $0.009$ kg water per kg ingredients during the conditioning process to $0.057$ kg water per kg ingredients (Fig. \ref{ED_fig5}c\&e). This suggests that, in this case, the lubrication layer thickness increases, filling more cavities in the asperities between the pellet and the metal die, transitioning into the mixed lubrication regime \cite{Wasche2014}. In the mixed regime, the contact between the two surfaces is reduced, resulting in less wear on the pellet's surface. 

\clearpage

\subsubsection{Mixed lubrication and the apparent coefficient of friction}
In the mixed lubrication regime, the apparent coefficient of friction, $\mu$ (-), is partially determined by the friction between the two solid surfaces, $\mu_{\mathrm{solid}}$ (-), and partially determined by the friction within the lubricating film, here water, $\mu_{\mathrm{water}}$ (-) \cite{Wasche2014}. The ratio between these two contributors changes with a change in the lubrication film thickness \cite{Taylor2022} (see Eq. \ref{equation_2}):

\begin{equation}\label{equation_2}
\mu = \mu_{\mathrm{solid}}X+\mu_{\mathrm{water}}(1-X)
\end{equation}

Here, $X$ (-), with values from $0$ and $1$, represents the ratio between the load carried by the solid-solid contact interface with respect to the total contact load. The remainder, $(1-X)$, represents the load carried by the fluid film between the two surfaces. In the case of boundary lubrication, $X$ is equal to 1, and the apparent coefficient of friction, $\mu$, is fully determined by the solid-solid sliding friction. Non-lubricated pellet extrusion would occur during pellet production without the addition of water, steam, fat, oil, or other fluid lubricants. However, in the presence of a water film between the two sliding surfaces, $X$ reduces to values smaller than $1$, and the apparent coefficient of friction, determined from the experimental processing data, decreases.

\subsubsection{The effect of production rate on the mechanical energy use}
\begin{figure}[!th]%
\centering
\includegraphics[width=0.5\textwidth]{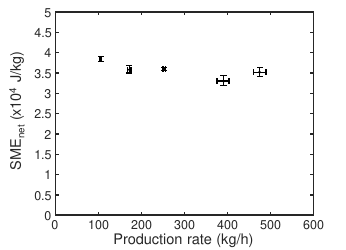}
\caption{The transition into the mixed lubrication regime is marked by a reduction in mechanical energy use ($SME_{\rm net}$) as the sliding velocity (or production rate) increases, corresponding to a decrease in the coefficient of friction, as observed. The error-bars correspond to the minimum and maximum values of two or three independent weighing samples taken during experimental runs.}
\label{fig:ED_Q_vs_SME}
\end{figure}

In the mixed lubrication regime, the coefficient of friction decreases with increasing sliding velocity, though the rate of this decrease depends on specific factors such as surface hardness, roughness, lubricant properties, and load~\cite{bongaerts_soft_tribo}. As discussed in previous sections, the amount of water added during the steam conditioning process determines whether pellet extrusion occurs in the boundary regime (with visible surface striations) or transitions into the mixed lubrication regime (where surface striations disappear). Although the operational space on the pellet press is limited compared to typical rheological devices, which can span several decades of shear rates, we were able to study the effect of production rate on mechanical energy use across a range of 100 to 500 kg/hr during pellet extrusion, with the addition of 0.057 kg of water per kg ingredients. As shown in Fig.  \ref{fig:ED_Q_vs_SME}, a slight decrease in mechanical energy use is observed as production rate increases. While multiple factors may contribute to this trend, one possible explanation is the reduced coefficient of friction due to higher sliding speeds in the mixed lubrication regime, consistent with typical Stribeck behavior. 

\subsection{Physical pellet quality} \label{ED physical quality}
\begin{figure*}[!th]%
\centering
\includegraphics[width=\textwidth]{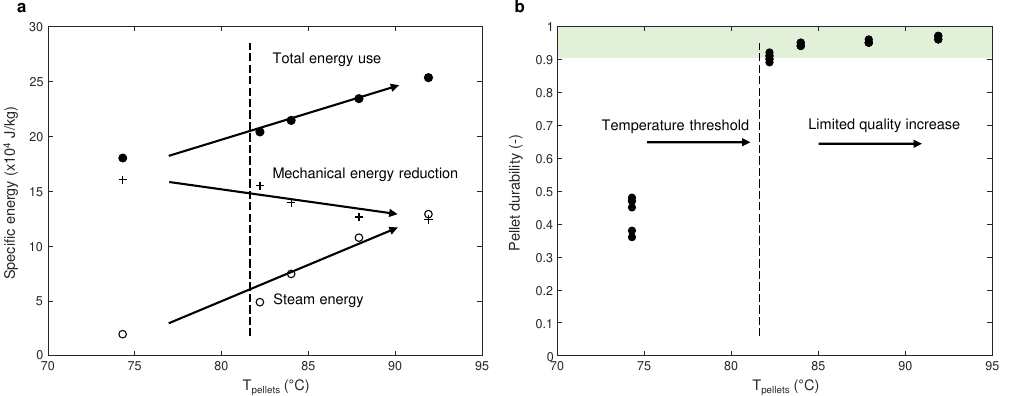}
\caption{(a) During pelleting, energy is required for the production of steam (marked by ``o"). In addition, electrical motor power is dissipated into the pellets (marked by ``+"). Here, the total mechanical energy dissipated by the press, including the idle load of the pellet press (gross mechanical energy), is represented. The sum of the steam energy and mechanical energy determines the total energy used in pellet production (marked by ``•"), which includes both conditioning and pelletizing. An optimum energy use exists, coupled with a normative pellet durability value (here $0.9$, equal to $90\%$), specifying the minimum amount of energy required to produce a high-quality pellet. This optimum is reached when pellet durability (b) no longer increases significantly with the invested energy during production (a). Below this optimum, pellets are still formed; however, the quality is insufficient for downstream processes, while above this optimum, the quality increases disproportionately with the energy investment (b). This pellet durability increase with increasing steam usage is consistent with \cite{Skoch1981}.}\label{ED_fig_PQ}
\end{figure*}
Throughout pellet production, physical pellet quality is continually monitored for quality control. The key metric for physical pellet quality is pellet durability, which refers to the pellet's ability to withstand handling and transportation without disintegrating \cite{Thomas1996}. High pellet durability indicates a strong resistance against breakage, which is critical in minimizing dust formation. Dust formation poses safety risks, including dust explosions, and health hazards associated with the inhalation of fine dust particles for both humans and animals. Therefore, achieving high durability is a crucial aspect of any pelleting process.

Producing a durable pellet with minimal energy use involves precise control over heat fluxes within the pellet production line. This control allows efficient adjustment of ingredient and pellet temperatures to facilitate bond formation and produce durable pellets (see Fig. \ref{ED_fig_PQ}b). The heating process begins with the addition of steam during steam conditioning, followed by heating through friction during pellet extrusion. Increasing steam usage enables a reduction in mechanical energy (see Fig. \ref{fig:results}b and Fig. \ref{ED_fig_PQ}a). However, it is essential to note that steam production itself is energy-intensive (see Fig. \ref{ED_fig_PQ}a), here estimated through assuming a $100\%$ efficiency of the steam-boiler. Therefore, finding a balance between steam and mechanical energy fluxes is crucial for optimizing pellet production for energy efficiency. Beyond a certain temperature threshold for a specified ingredient composition, further increases in total energy input (see Fig. \ref{ED_fig_PQ}a - "Total energy use") yield no significant durability improvements (see Fig. \ref{ED_fig_PQ}b). Above this optimum process temperature, production costs increase significantly with continued steam usage. To mitigate these costs, replacing part of the added steam with tap water during conditioning might be considered. Tap water can contribute to lubrication, reducing mechanical energy use without the energy costs associated with an increased steam production. Therefore active monitoring of the lubrication layer, the mechanical energy use and product temperatures, is a requirement for future process optimization during pellet extrusion.

\clearpage

\section{Methods \& Protocols}\label{sec:methods}

\subsection{Materials}
A total of 18,000 kg of milled material was obtained from Research Diet Services (Wijk bij Duurstede, The Netherlands). The material was ground using a hammer mill equipped with 4-mm screens. The material, a 50/50\% w/w mixture of cleaned maize kernels and low-sugar sugar beet pulp, was delivered in two separate orders: the first batch (\textit{Batch 1}) contained 7,000 kg, and the second batch (\textit{Batch 2}) contained 11,000 kg. Batch 1 was used to obtain the results presented in all figures except Fig. \ref{ED_fig_LD-ratio}, Batch 2 was used solely for the results presented in Fig. \ref{ED_fig_LD-ratio}. The products, being of organic/natural origin, may differ in their chemical compositional between Batch 1 and Batch 2. However, these variations were not the focus of this study, therefore the chemical composition was not analyzed.

\subsection{Pellet production and process monitoring}
A RMP200 ring-die pelletizer (Münch Edelstahl GmbH, Hilden Germany) served as the pilot-scale pellet extrusion system. Configured with two rollers, the ring-die featured 6 mm diameter die-holes and a compression ratio (L/D) of 12, corresponding to a channel length of 72 mm. In addition to the 6 mm L/D 12 die, three additional dies with 6 mm diameter holes were used for the results presented in Fig. \ref{ED_fig_LD-ratio}, with L/D ratio's 6, 9.3 and 16 respectively. The roller-die gap was set at 0.20 mm, and the die rotation frequency maintained at 5 Hz.

Steam injection and mixing, facilitated by a single-screw steam conditioner (Münch Edelstahl GmbH, Hilden Germany), adjusted the water content of the ingredient mixture prior to pelletization. Super-heated steam was injected at 140°C and used to set the ingredient temperature inside the steam conditioner.

Each run was initialized with a 20 minutes period, during which feeder speed and steam injection rate were adjusted to obtain steady-state. This steady-state was confirmed by monitoring of the electrical current use (A, ampere) of the pellet press using a Janitza Power Analyzer UMG-604 (Janitza, Lahnau, Germany).

Registered process parameters include ingredient mixture temperature and moisture content before and after steam conditioning, pellet temperature and moisture content after pellet extrusion, electrical current, and production capacity (kg/h). From which the change in moisture content ($\Delta \mathrm{mc}$, in kg water per kg ingredients) during the steam conditioning process, the change in temperature over the extrusion die ($\Delta\mathrm{T_{die}}$) and the specific mechanical energy (SME) were calculated. The temperature measurements were performed as described in \cite{Bastiaansen2023}, the moisture content was determined as described in \cite{Vego2023} and the production capacity (also: mass-flow-rate) was determined as described in  \cite{Bastiaansen2023}, using a 2 minute measurement period.

The gross and net SME (kWh/t or J/kg) were determined according to the method described in \cite{Bastiaansen2023}, where the gross SME is calculated using the full-load current and the net SME is calculated by subtracting the idle-load current from the the full-load current. This enables us to differentiate between the total amount of energy dissipated (idle-load plus pellet extrusion) by the press and the amount of energy dissipated in the pellet extrusion process only.

During the processing of Batch 2, additionally, the steam-flux (kg steam per kg ingredients) was recorded (H250, Krohne, Duisburg, Germany), instead of the change in moisture content before and after conditioning.

During the processing of Batch 1, the effect of steam dosing on pellet extrusion was evaluated by conducting five steady-state runs in random order at constant production capacity (approx. 250 kg/h) and variable steam injection rate, thereby tuning $\Delta \mathrm{mc}$ as specified in Fig. \ref{fig:main}d. During the processing of Batch 2, the effect of L/D ratio and steam dosage on the specific mechanical energy use was evaluated using a constant production capacity (approx. 250 kg/h) and at two levels of $\Delta \mathrm{mc}$ (0.035 and 0.053 kg steam per kg ingredients, respectively), while changing the die-configuration as specified in Fig. \ref{ED_fig_LD-ratio}b.

\subsection{Drying setup and analysis}
The lubrication layer thickness was determined according to the method described in~\cite{Benders2022}, using a model drying setup. The drying setup, consists of a 3D-printed air-tunnel, into which a controlled air-flow (40 L/min, FL-2044, Omega Engineering Limited, Manchester, United Kingdom) is supplied from one side. The air passes through a 3D printed (PolyLite ASA - black, Polymaker) sample container containing a sample of pellets, which is placed on top of an analytical balance (ME204, Mettler-Toledo B.V., Tiel, the Netherlands). During the drying process, the temperature and relative humidity (RH) of the air entering and exiting the sample container with pellets over time was recorded using  a TSP01 RevB sensor (Thorlabs Inc.).

The sample container (6x6x5 cm) was equipped with two sheets of stainless steel woven wire mesh (6x6 cm) with aperture of 2.4 mm, positioned at opposite sides of the container, allowing the air to pass through the filled sample container, whilst retaining the pellets. The input airline was connected to a compressed air line, available in the process hall. The flow rate of 40 L/min, corresponds to an air velocity of 0.19 m/s upon entering the sample container. Samples of approximately 100g of fresh pellets were collected in the sample container directly below the pellet press. The mass of the sample container was recorded for a period of 15 minutes at a sampling frequency of 1 Hz, after which the moisture ratio ($\mathrm{MR}$, see Eq. \ref{appB_eq3}) was determined. The air dried material was transferred to a plastic bag and sealed immediately. The moisture content (kg water per kg pellet) of the pellets after drying was subsequently determined according to the method described in \cite{Vego2023}.

Model drying experiments were performed three times (20 minutes apart) during the steady-state of the pelleting press by placement container with freshly sampled pellets in the drying setup. The saturation time period, $t_{\mathrm{sat}}$, was determined visually from the RH-curves, and is defined as the time period during which the RH-value is larger than 95\%, indicated by the green area in Fig. \ref{fig:main}b. The amount of evaporated water (kg water per kg pellet) was determined through multiplication of the averaged moisture ratio ($\mathrm{MR}$, unitless) over time and the averaged change in moisture content (kg water per kg pellet). The latter is determined as the difference in the moisture content at the start ($\mathrm{mc}$ right after pellet extrusion) and at the end of the drying experiment ($\mathrm{mc}$ of the pellets collected, bagged and sealed).

\subsection{Calculating $\lambda$}
The lubrication layer thickness is calculated from the amount of evaporated water at time $t_{\mathrm{sat}}$ during the drying experiment \cite{Benders2022}. The amount of evaporated water ($X$, kg water per kg pellet) is converted to an annular volume of water ($V_{\mathrm{water}}$) on top of the pellets surface, with $V_{\mathrm{water}}=\pi l (R^2-r^2)$. The lubrication layer thickness $\lambda$ is calculated using $\lambda = R - r$, in which $R$ is the outer radius of the annulus (water layer plus pellet radius) and $r$ the radius of a pellet. $R$ is unknown but is determined from $R^2=(V_{\mathrm{water}}/V_{\mathrm{pellet}}+1)r^2$ using $V_{\mathrm{water}}/V_{\mathrm{pellet}} = X/\rho_{\mathrm{water}}\times\rho_{\mathrm{pellet}}$. The latter equation converts the amount of evaporated water in kg water per kg pellet, into the evaporated water volume per volume pellet. To calculate $R$, the density of the pellet, $\rho_{\mathrm{pellet}}$ is assumed to be $1200\ \mathrm{kg_{\mathrm{pellet}}\ m^{-3}}$\cite{Stelte2011}, the radius of the a pellet is approximated by the inner radius of the extrusion channel ($D/2$) here $3\ \mathrm{mm}$ (see Fig. \ref{ED_fig2}), and the density of water $\rho_{\mathrm{water}}$ is approximated at $1000\ \mathrm{kg_{\mathrm{water}}\ m^{-3}}$.

\subsection{Rehydration of pellets}
As a control drying experiments, a sample of dried pellets were rehydrated through hygroscopic vapor sorption. This sample of approximately 300 grams of pellets was stored in a container maintained at \textgreater95\% RH for over 1 week, before being dried in the model dryer, using the same drying conditions as during the drying of fresh pellets. Two drying runs were performed, one lasting 3.5 hours and the other approximately 15 hours (yellow lines in Fig. \ref{ED_fig1}c). 

\subsection{Physical pellet quality}
Pellet durability was determined using a Holmen NHP100 (Tekpro, USA) tester with a 3-mm screen. During each steady-state run, three pellet samples of approximately 2 kg were sampled directly after leaving the press. The sampling was performed approximately 20 minutes apart. The 2 kg samples were subsequently cooled at ambient air for 10 min and transferred into two plastic pots (1 L). Approximately 100 grams of pellets, from each pot, underwent testing in the rotating drum mechanism, simulating mechanical stress \cite{Thomas1996}. The Pellet Durability Index (PDI) was calculated based on weight loss after tumbling, indicating the percentage of intact pellets after testing. A high pellet durability (\%), indicates a high resistivity of pellets against abrasion during handling and transport.

\subsection{Synchrotron-based X-ray microtomography}
Computed tomography (CT) scans were conducted on pellets produced under various process conditions utilizing the ANATOMIX beamline at the synchrotron SOLEIL. Each pellet was securely mounted in custom 3D printed sample holders, allowing for individual scanning. These holders were positioned in the Huber goniometer head, maintaining a consistent distance of 20 cm from the scintillator. Image acquisition utilized a 2048x2048 pixel detector with a pixel size of 3.07 µm, such that the pellet (approximately 6 mm in diameter) fits entirely within the detectors field of view. The scanning process involved a rotation step of 0.09°, resulting in 2000 projections collected over a 180° rotation. Reconstruction of the acquired data into a 3D image volume was performed using a Paganin filter. In Fig. \ref{ED_fig4}, four scans are shown of four individual pellets produced at two steam injection levels (2x2).

\subsection{Density profile}
The radial density profile of an X-ray tomogram was determined using a custom MATLAB (R2023a) code. The 3D image volume of a pellet was segmented into 30 equidistant slices along the axial direction (top to bottom), with each slice processed individually.

Initially, the center of the pellet was identified, and subsequently, an Euclidean distance map was generated. This map enabled the calculation of the distance of each pixel in the slice relative to the pellet's center point, measured in pixels. Then, the intensity values of each pixel at a given distance (1-pixel bins) were averaged. This process was repeated for all 30 slices.

To ensure data quality, the average intensity of each image slice was utilized as a threshold to discard slices with low X-ray photon counts, resulting from drift in the alignment of the monochromatic beam with respect to the sample and the detector. The remaining slices were then averaged to obtain the final average density profile along the radial coordinate, representing the distance from the pellet's center point.

\subsection{Scanning electron microscopy}
SEM was performed on a FEI Magellan 400 (FEI Electron Optics B.V., The Netherlands) using an acceleration voltage of 2.0 kV. The pellets were mounted on aluminum stubs with double-sided adhesive, conductive carbon tape. Pellets were cut along the longitudinal axis. The outer and inner pellet surface was oriented upward and sputter-coated with 12 nm of Tungsten under a 45° angle (Leica EM SCD500, Leica Microsystems, The Netherlands). 

\subsection{Surface characterisation}
An automatic, noncontact laser confocal microscope (VK-X1000 Series; KEYENCE, Japan) was used to analyze surface topology of pellets at ×20 magnification with a corresponding pixel size of 0.686 $\mathrm{\mu m}$. The surface height profile was exported in the KEYENCE software and further analyzed using a custom code in MATLAB (R2020b). In MATLAB the height data was treated as image data (768x1024 pixels). A 3D surface profile was generated by converting the height data into a stereolithography (.STL) file. The background curvature, the cylinder shape of the pellets, was determined and afterwards subtracted from the original height data by using a Gaussian filter with a large value for sigma. The value for sigma (= 60) was chosen to smooth the height data over a large area, which prevents smoothing the much smaller features at the pellets surface. The flattened height data was then trimmed by 60 pixels along all four edges to remove artifacts at the image boundaries (648x904 pixels). The trimmed and flattened height data was used for visualization of the height profiles (see Fig. \ref{ED_fig5}). 

\clearpage
\section{appendix}
\subsection{Simulation details COMSOL}\label{appendix:COLSOM_Model}

Numerical simulations were performed using COMSOL Multiphysics software to analyze the drying behavior of pellets. This study focused on the impact of the redistribution of water along the radial coordinate due to radial phase migration. The objective of these simulations was to qualitatively compare the experimental drying data with the potential impact, as revealed through numerical modeling, of radial phase migration on drying behavior. It is essential to note that the experimental data were obtained through the drying of a container with approximately 100 grams of sample material, whereas the numerical effect of radial redistribution of moisture was simulated on a single pellet level. Evaporation from the top and bottom ends of a pellet was ignored. Therefore, drying of a pellet was simulated along a 1D line of length $r_p$ ($3\times10^{-3}\ \mathrm{m}$), the radius of a pellet produced in a 6 mm pellet die, with a symmetry axis at $r = 0$.

The governing partial differential equation for the transport of moisture, $M$, throughout the pellet was based on Fick's law:
\begin{equation}\label{appB_eq1}
\frac{\partial M}{\partial t} = \frac{1}{r} \frac{\partial}{\partial r}\left( r D_{\rm eff}\frac{\partial M}{\partial r}\right)
\end{equation}
where the rate of moisture loss $\partial M/\partial t$ is controlled by the moisture diffusivity inside the pellet, $D_{\rm eff}$.

To solve eq. \ref{appB_eq1}, a convective boundary condition was used, formulated as an analogue to the Hertz-Knudsel-Schrage equation, wherein the gradient in vapor pressure drives the evaporation process, as expressed by equation \ref{appB_eq2}:
\begin{equation}\label{appB_eq2}
-D_{\rm eff}\left .\frac{\partial M}{\partial r}\right|_{r=r_p}= k_c\left(0 - M \right)
\end{equation}
where $k_c$ is the mass transfer coefficient. The concentration outside of the pellet was set to 0 based on the following rationale: During pellet production, water is added to the ingredient mixture through the condensation of steam during steam conditioning. This added water elevates the moisture content within the pellets above its equilibrium moisture content ($M\geq M_\mathrm{eq}$), the moisture content of the ingredient mixture pre-conditioning. The added water will evaporate until the equilibrium moisture content is restored ($M= M_\mathrm{eq}$). This equilibrium state implies that there is no longer a gradient in partial vapor pressure between the air and the ingredient mixture, resulting in equal rates of evaporation and condensation. Consequently, the effective external moisture content was set to 0 (as per equation \ref{appB_eq2}), indicating that all added water will eventually evaporate as the sign of $M$ is positive (reflecting the addition of water during pellet production).

All simulations were initialized with an initial moisture content of $+2.5\%$ (kg water per kg material), a typical increase in moisture content during steam conditioning.

The value for $D_{\rm eff}$ was estimated at $5\times10^{-11}\ \mathrm{m^2\ s^{-1}}$ which matches typical values reported in literature for foodstuffs \cite{Panagiotou2004} and closely matches the experimental drying rate of the uniformly rehydrated pellets shown by the yellow lines in Fig. \ref{ED_fig1}. The value of $k_c\ (2\times 10^{-2}\ \mathrm{m\ s^{-1})}$ was selected such that the mass Biot number $\mathrm{Bi_m} = \frac{k_c\times r_p}{D_{\rm eff}} = 1.2\times10^{6} \gg 1$. Therefore, the rate of diffusion within the pellet is the evaporation rate-limiting factor, as can be expected for hygroscopic organic material that slowly absorbs or releases water depending on the relative humidity of the air in the environment \cite{Vego2023}.

The normalized moisture ratio, $\mathrm{MR}$, shown in Fig. \ref{ED_fig1}b, was obtained by integrating the moisture content, $\int M r\ dr$, over the domain $r = 0$ to $r = r_p$, to determine the amount of water, $X$, at time $t$, at $t=0 \mathrm{s}$, and at $t=900\mathrm{s}$:
\begin{equation}\label{appB_eq3}
\mathrm{MR} = \frac{X(t)-X(t=900\mathrm{s})}{X(t=0\mathrm{s})-X(t=900\mathrm{s})}
\end{equation}.
\subsection{Energy dissipation in a liquid film }\label{appendix:EnergyDissipation}
The surface topology scans, exhibiting clear striation patterns upon the addition of minimal water during pellet manufacturing and the disappearance thereof with increased water addition, suggest that we are within the mixed lubrication regime. In this regime, where $X$ ranges between $1$ and $0$, the work of friction is predominantly determined by solid-solid contact friction (see eq. \ref{equation_2}). The exact contributions of $\mu_{\mathrm{solid}}$ and $\mu_\mathrm{water}$ at operational processes are unknown. However, the contribution of friction in the water film can be estimated by assuming a full separation of the two sliding surfaces, in which case all friction is assumed to result from shearing of the lubrication layer. Through careful estimates, we can determine the specific energy dissipation due to the shearing of the water film (see eq. \ref{equation_last}):
\begin{equation}\label{equation_last}
    \frac{P_\mathrm{diss}}{Q_\mathrm{m}} = \frac{V_\mathrm{water}\eta_{\mathrm{water}}\dot\gamma_{\mathrm{water}}^2}{\dot V_\mathrm{pellet}\rho_{\mathrm{pellet}}}
\end{equation}
Here, $P_\mathrm{diss}$ is the energy dissipation in the water film, $Q_\mathrm{m}$ is the mass-flow-rate of the pellet through the die-hole. $V_\mathrm{water}$ is the volume of the sheared water, an annulus with outer radius $r$ (here $3\ \mathrm{mm}$) and inner radius $r-\lambda$, using the height of the water film, $\lambda$ (here approximated by $10\ \mathrm{\mu m}$) and die-hole channel length $L$ (here $72\ \mathrm{mm}$). The viscosity of water, $\eta_{\mathrm{water}}$, at room temperature is approximately $1\ \mathrm{mPa\ s}$. The shear rate $\dot\gamma_{\mathrm{water}}$ is calculated from the extrusion velocity, $v$ (approximately $1\ \mathrm{ms^{-1}}$), the velocity at which the pellet moves through the die when the roller passes over the die-hole, and the height of the water film, $\lambda$. Thus, the shear rate is approximately $10^5\ \mathrm{s^{-1}}$. $\dot V_\mathrm{pellet}$ is the volumetric flow rate of a cylindrical pellet through the die-hole, for a pellet with radius $r-\lambda$ and using a displacement rate equal to the extrusion velocity ($\dot V_\mathrm{pellet} = \pi (r-\lambda)^2 v$). And the pellet's density, $\rho_{\mathrm{pellet}}$ (here approximated at $10^3\ \mathrm{kg\ m^{-3}}$).

The specific energy dissipation, in the case where all the shear stresses would act on a water-based lubrication layer of $10\ \mathrm{\mu m}$ thickness, is estimated to be $5\ \mathrm{J\ kg^{-1}}$. However, the experimental data shows an net energy dissipation of more than $60\ \mathrm{kJ\ kg^{-1}}$ upon the presence of a $10\ \mathrm{\mu m}$ lubrication layer during the pelleting trial (see Fig. \ref{fig:results}). Therefore, we conclude that we are in the mixed lubrication regime, in which the predominant part of the apparent friction is determined by the solid-solid sliding friction, $\mu_{\mathrm{solid}}$, and not by the friction within the water film, $\mu_{\mathrm{water}}$, see eq. \ref{equation_2}.

\bibliography{apssamp}
\end{document}